\documentclass[12pt]{amsart}          %
\usepackage{amsfonts} 
\usepackage{epsfig}
\usepackage{hhline,eufrak,euscript,delarray}

\author{ Vlady RAVELOMANANA }
\email{vlad@lipn.univ-paris13.fr}
\address{Vlady RAVELOMANANA, LIPN -- UMR 7030, Institut Galil\'ee --Universit\'e de Paris-Nord,
99, Avenue J. B. Cl\'ement. F 93430 Villetaneuse, France.}
\title[]{Another proof of Wright's inequalities}

\def\Vz{\vartheta_z}

\def\SUP#1{\overline{W}_{#1}}
\def\INF#1{\, \underline{W}_{\,  #1}}

\def\INFPSI{{\Psi}}

\def \beq{\begin{equation}}
\def \eeq{\end{equation}}
\def \be{\begin{eqnarray}}
\def \ee{\end{eqnarray}}
\def \ben{\begin{eqnarray}}
\def \een{\end{eqnarray}}
\def \l{\left}
\def \r{\right}
\def\coeff#1{\left[ #1 \right]}

\begin{document}
\label{firstpage}

\begin{abstract}
We present a short way of proving the 
inequalities obtained by Wright in 
$\left[\mbox{{\em Journal of Graph Theory}, 4: 393 -- 407 (1980)}\right]$
 concerning the number of 
 connected graphs with $\ell$ edges more than vertices.
\end{abstract}

\maketitle

\section{Preliminaries}

For $n\geq 0$ and $-1\leq \ell \leq {n \choose 2} -n$, let $c(n, \, n+\ell)$ be the 
number of connected graphs with $n$ vertices and $n+\ell$ edges. Quantifying
$c(n, \, n+\ell)$ represents one of the fundamental tasks
 in the theory of random graphs.
It has been extensively studied  since the Erd\H{o}s-R\'enyi's paper \cite{ER59}.
 The generating functions
associated to the numbers $c(n, \, n+\ell)$ are due to Sir E.~M.~Wright in a series of
papers including \cite{Wr77,Wr80}. He also obtained the asymptotic formula
for $c(n,\, n+\ell)$ for every $\ell = o(n^{1/3})$. Using different methods,
 Bender, Canfield and McKay  \cite{BCM90}, 
Pittel and Wormald \cite{PITTEL-WORMALD} and  van der Hofstad and Spencer
\cite{Sp+} were able to determine the asymptotic value of $c(n, \, n+\ell)$
for all ranges of $n$ and $\ell$.

For $\ell \geq -1$,
let $W_{\ell}$ be the exponential generating function (EGF, for short) 
of the family of connected graphs
with $n$ vertices and $n+\ell$ edges. Thus,
$W_{\ell}(z) = \sum_{n=0}^{\infty} c(n, \, n+\ell) \frac{z^n}{n!}$.
Let $T(z)$
be the EGF of the Cayley's rooted labeled trees. It is well known that
$T(z) = z \, e^{T(z)} = \sum_{n\geq 1} n^{n-1} \frac{z^n}{n!}$
(see for example \cite{FS+,JKLP93}).
Among other results, Wright proved that
 the functions $W_{\ell}(z)$, $\ell \geq -1$, can be expressed in terms of 
 $T(z)$. Such results allowed penetrating and precise analysis
when studying random graphs processes as it has been shown
for example in the giant paper \cite{JKLP93}.
Throughout the rest of this note, all formal power series are
univariate. Therefore, for sake of simplicity we will often omit
the variable $z$ so that $T \equiv T(z)$, $W_i \equiv W_i(z)$ and so on.

We need the following notations.

\noindent \textbf{Definition.} If $A$ and $B$ are two formal power series such that
for all $n \geq 0$ we have $\coeff{z^n} A(z) \leq \coeff{z^n}B(z)$ then
we denote this relation $A \preceq B$ or $A(z) \preceq B(z)$.

The aim of this note is to provide an alternative and 
generating function based proof of the 
inequalities obtained by Sir Wright in \cite{Wr80} (in particular,
he used numerous intermediate lemmas). More precisely,
Wright obtained the following.

\noindent \textbf{Theorem (Wright 1980).} 
\textit{
Let $b_1=\frac{5}{24}$ and $c_1=\frac{19}{24}$.
Define recursively $b_{\ell}$ and 
$c_{\ell}$ by
\begin{equation}
2 (\ell+1) b_{\ell+1} = 3\ell(\ell+1)b_{\ell}+ 3 \sum_{t=1}^{\ell-1} t(\ell-t)b_t b_{\ell-t} \, ,
\qquad (\ell \geq 1)
\label{eq:B_K}
\end{equation}
and
\begin{eqnarray}
 2 (3\ell+2) c_{\ell+1} &=& 8 (\ell+1) b_{\ell+1} + 3\ell b_{\ell} + (3\ell+2)(3\ell-1) c_{\ell} \cr
 & + & 6\sum_{t=1}^{\ell-1} t (3\ell - 3t -1) b_t c_{\ell-t}\, , \qquad (\ell \geq 1) \quad .
\label{eq:C_K}
\end{eqnarray}
Then, for all $\ell \geq 1$
\beq
\frac{b_{\ell}}{(1-T(z))^{3\ell}} - \frac{c_{\ell}}{(1-T(z))^{3\ell-1}} 
\preceq W_{\ell}(z) \preceq \frac{b_{\ell}}{(1-T(z))^{3\ell}} \quad .
\label{eq:MAIN}
\eeq}
(\ref{eq:MAIN}) is known as \textit{Wright's inequalities} and such results 
 has been extremely useful in the enumerative study of graphs as well as in
the theory of random graphs \cite{Bollobas,JKLP93,JLR2000,LUCZAK-BOUNDS,Myself}.

Our proof of (\ref{eq:MAIN}) is based upon two ingredients:

\bigskip

\noindent
\textbf{Fact 1.} We know that the EGFs $W_{\ell}$ satisfy 
$W_{-1} = T - \frac{T^2}{2}$, $W_{0} = - \frac{1}{2} \log{(1-T)} - \frac{T}{2} - \frac{T^2}{4}$ 
 and
\beq
\left( 1 - T \right) \Vz W_{\ell+1} + \left(\ell+1\right) \,  W_{\ell+1}  = %
\left( \frac{\Vz^2 - 3 \Vz}{2} - \ell \right)W_{\ell} +
\frac{1}{2} \sum_{k=0}^{\ell} \left( \Vz W_{k} \right) \left( \Vz W_{\ell -k} \right) \,,
\quad (\ell \geq 0) \,,
\label{eq:FUNCTIONAL}
\eeq
where $T=T(z)$, $W_k = W_k(z)$ and 
$\Vz = z \frac{\partial}{\partial z}$ corresponds to marking a vertex
(such combinatorial operator consists to choose a vertex among the others).
For the combinatorial sense of (\ref{eq:FUNCTIONAL}), we refer the reader
to \cite{BCM90,JKLP93} or \cite{Wr77}.

\bigskip

\noindent
\textbf{Fact 2.}
Let $A$ and $B$ be two formal power series and $\ell \in \mathbb{N}$.
If $(1-T) \, \Vz A  + (\ell+1)\, A \preceq (1-T) \, \Vz B  + (\ell+1)\, B$
then $A \preceq B$.

\noindent
To prove Fact 2, fix $\ell \geq 0$. We write
\beq
B(z)-A(z) = \sum_{n=0}^{\infty} (b_n - a_n) \frac{z^n}{n!} %
\quad \mbox{ and } \quad \forall n, \, c_n = b_n - a_n \, .
\eeq
Suppose that
 $(1-T) \, \Vz A  + (\ell+1)\, A \preceq (1-T) \, \Vz B  + (\ell+1)\, B$.
We then have
\ben
& & n! \coeff{z^n} \l( \l(1-T(z)\r) \Vz \l(B(z) - A(z)\r)%
 + (\ell+1)\l(B(z)-A(z)\r) \r)  = \cr
& & \qquad \qquad %
(n+\ell+1) c_n - \sum_{k=1}^{n} {n \choose k} k^{k-1} (n-k) c_{n-k} \geq 0\, .
\een
It is now easily seen that $\forall n, c_n \geq 0$. Therefore, $A \preceq B$.

\bigskip

Our proof of (\ref{eq:MAIN}) is divided into two parts each of each are given in the
next Sections. 
\section{Proof of $W_{\ell} \preceq \frac{b_{\ell}}{(1-T)^{3\ell}}$}
Define the family $(\SUP{\ell})_{\ell \geq 0}$
as 
$\SUP{0} = - \frac{1}{2} \log{(1-T)}$
and for $\ell \in \mathbb{N}^{\star}$,
 $\SUP{\ell} = \frac{b_{\ell}}{(1-T)^{3\ell}}$.
Observe that we have $W_0 \preceq \SUP{0}$ and 
  $W_1 \preceq \SUP{1}$ has been proved in \cite{Wr80}. Now, we can proceed by induction.
Suppose that for $2 \leq i \leq \ell$, 
$W_i \preceq \SUP{i} = \frac{b_i}{\left(1-T\right)^{3i}}$ 
and let us prove
that $W_{\ell+1} \preceq \SUP{\ell+1} = \frac{b_{\ell+1}}{(1-T)^{3\ell+3}}$.
Simple calculations show that
\beq \label{SHOW1}
\l(\frac{\Vz^2 - \Vz}{2}\r) \SUP{\ell} \; \preceq \; \frac{\Vz^2}{2} \SUP{\ell}
\; \preceq \; \frac{3\ell(3\ell+2)}{2} \, \frac{b_{\ell}}{(1-T)^{3\ell+4}}
- \frac{3\ell(3\ell+2)}{2} \, \frac{b_{\ell}}{(1-T)^{3\ell+3}} \, ,
\eeq
\beq \label{SHOW2}
\l(\Vz \SUP{0}\r) \; \l(\Vz \SUP{\ell}\r) \; \preceq \; %
\frac{3\ell b_{\ell}}{2} \,  \frac{b_{\ell}}{(1-T)^{3\ell+4}} 
- \frac{3\ell b_{\ell}}{2} \,  \frac{b_{\ell}}{(1-T)^{3\ell+3}} %
\quad \mbox{ and}
\eeq
\beq \label{SHOW3}
\frac{1}{2} \sum_{p=1}^{\ell-1} \l(\Vz \SUP{p}\r) \l(\Vz \SUP{\ell-p}\r) %
\; \preceq \; \frac{1}{2} %
\l( \sum_{p=1}^{\ell-1} 9p(\ell-p)b_p b_{\ell-p} \r) \;
\l( \frac{1}{(1-T)^{3\ell+4}} - \frac{1}{(1-T)^{3\ell+3}}  \r) \, .
\eeq
Summing (\ref{SHOW1}), (\ref{SHOW2}), (\ref{SHOW3}),
 using the recurrence (\ref{eq:B_K}) and the induction hypothesis,
 we find that
\beq
(1-T)\Vz W_{\ell+1} + (\ell+1)W_{\ell+1} %
\preceq \frac{3(\ell+1)b_{\ell+1}}{(1-T)^{3\ell+4}} 
-  \frac{3(\ell+1)b_{\ell+1}}{(1-T)^{3\ell+3}} \, . 
\eeq
Since
\beq
(1-T)\Vz \SUP{\ell+1} + (\ell+1)\SUP{\ell+1} %
= \frac{3(\ell+1)b_{\ell+1}}{(1-T)^{3\ell+4}} 
-  \frac{2(\ell+1)b_{\ell+1}}{(1-T)^{3\ell+3}} \, 
\eeq
by Fact 2, we have $\SUP{\ell+1} \succeq W_{\ell+1}$.

\section{Proof of $\frac{b_{\ell}}{(1-T)^{3\ell}} - \frac{c_{\ell}}{(1-T)^{3\ell-1}} \preceq W_{\ell}$}
Define  
$\INF{0} = W_{0}$ and for $\ell \in \mathbb{N}^{\star}$, 
$\INF{\ell} =  \frac{b_{\ell}}{(1-T)^{3\ell}} - \frac{c_{\ell}}{(1-T)^{3\ell-1}}$.
As before, we shall proceed by induction. We have $\INF{0} \preceq W_0$ and
\beq
W_1 - \INF{1} = \frac{13}{12 \, (1-T)} - \frac{1}{2} - \frac{T}{8} + \frac{T^2}{24} \, %
\succeq \frac{13}{12} \left( \frac{1}{(1-T)} - T - 1 \right) = %
\frac{13\, T^2}{12 (1-T)} \succeq 0 \, . 
\eeq
Suppose that for $2 \leq k \leq \ell$, 
$\INF{k} = \frac{b_{k}}{(1-T)^{3k}} - \frac{c_{k}}{(1-T)^{3k-1}} \preceq W_{k}$. 
We have to prove that 
$\INF{\ell+1} = \frac{b_{\ell+1}}{(1-T)^{3\ell+3}} - \frac{c_{\ell+1}}{(1-T)^{3\ell+2}} \preceq W_{\ell+1}$.
For this purpose, define ${\INFPSI}_{\ell+1}$ as 
\ben
& & {\INFPSI}_{\ell+1} = 
\left(\frac{\Vz^2-3\Vz}{2}-\ell\right)\, \INF{\ell} +
\left( \Vz\INF{0} \right) \left( \Vz\INF{\ell} \right)  
  +  \frac{1}{2} \sum_{k=1}^{\ell-1} \left( \Vz \INF{k} - %
\frac{(3\ell-1) c_{\ell} }{(1-T)^{3\ell}} \right) %
\left( \Vz \INF{\ell -k} \right) \cr
& & \qquad - \qquad %
\left( \frac{\alpha_{\ell}}{(1-T)^{3\ell+2}} +  \frac{\beta_{\ell}}{(1-T)^{3\ell+1}} %
+  \frac{\gamma_{\ell}}{(1-T)^{3\ell}} + \frac{\delta_{\ell}}{(1-T)^{3\ell-1}}\right)  \, , \cr
& &  
\label{eq:DEF_INFPSI}
\een
where $\alpha_{\ell}$, $\beta_{\ell}$, $\gamma_{\ell}$ and $\delta_{\ell}$ are given by
\ben
\alpha_{\ell} &=&  \frac{(7\ell+4)c_{\ell+1}}{2} - 3(\ell+1)b_{\ell+1} -%
 \frac{3}{4}\ell b_{\ell} + \frac{(3\ell-1)(3\ell+4)}{4} c_{\ell} \cr
 &+& \frac{1}{2} \sum_{t=1}^{\ell-1} (3t-1) c_{t} (3\ell-3t -1) c_{\ell-t} \,  ,
\label{eq:ALPHA}
\een
\ben
\beta_{\ell} & = &  - \frac{(3\ell+2)c_{\ell+1}}{2} + 2(\ell+1)b_{\ell+1}
 - \frac{3}{4}\ell b_{\ell} - \frac{(3\ell-1)(3\ell+4)}{4} c_{\ell} \cr
 &-& \frac{1}{2} \sum_{t=1}^{\ell-1} (3t-1) c_{t} (3\ell-3t -1) c_{\ell-t} \,  ,
\label{eq:BETA}
\een
\ben
\gamma_{\ell} =  \frac{ \ell b_{\ell}} {2} + \frac{ (3\ell-1) c_{\ell} }{2} \, \quad
\mbox{ and } \quad  \delta_{\ell} = - \frac{\ell-1}{2} c_{\ell} \quad .
\label{eq:GAMMA_DELTA}
\een
Rewritting the formal power series 
$\frac{\alpha_{\ell}}{(1-T)^{3\ell+2}} +  \frac{\beta_{\ell}}{(1-T)^{3\ell+1}}
 +  \frac{\gamma_{\ell}}{(1-T)^{3\ell}} + \frac{\delta_{\ell}}{(1-T)^{3\ell-1}}$
as follows
\ben
& & \frac{(7\ell+4)/2 \, c_{\ell+1} - 3 (\ell+1)b_{\ell+1}- 3/4 \ell b_{\ell} }{(1-T)^{3\ell+2}} %
- \frac{(3\ell+2)/2 \, c_{\ell+1} - 2(\ell+1) b_{\ell+1} + 3/4 \ell b_{\ell} }{(1-T)^{3\ell+1}} \cr
& & \qquad + \quad (3\ell-1)(3\ell+4)c_{\ell}\left(\frac{1}{(1-T)^{3\ell+2}}- \frac{1}{(1-T)^{3\ell+1}}\right) \cr
& & \qquad + \quad \frac{2\ell b_{\ell}}{2 (1-T)^{3\ell}} + %
\left( \frac{(3\ell-1)c_{\ell}}{2(1-T)^{3\ell}} - %
\frac{(\ell-1)c_{\ell}}{2(1-T)^{3\ell-1}} \right) \, ,
\een
it is easily seen that if the quantity (coming from the denominators of
the $2$ first terms of the above equation)
\ben
(2\ell+1) c_{\ell+1} - (\ell+1) b_{\ell+1} - \frac{3}{2} \ell b_{\ell} \geq 0
\label{eq:FACILE}
\een
then
 $\frac{\alpha_{\ell}}{(1-T)^{3\ell+2}} +  \frac{\beta_{\ell}}{(1-T)^{3\ell+1}}
 +  \frac{\gamma_{\ell}}{(1-T)^{3\ell}} + \frac{\delta_{\ell}}{(1-T)^{3\ell-1}} \succeq 0$. 
(We used $1/(1-T)^a \succeq 1/(1-T)^b$ if $a\geq b$).

\noindent
Using (\ref{eq:B_K}) and (\ref{eq:C_K}), after simple algebra we have (\ref{eq:FACILE}).
Therefore by construction, %
$\mbox{RHS of} (\ref{eq:FUNCTIONAL}) \succeq {\INFPSI}_{\ell+1}$. 
 After nice cancellations,  it yields
\beq
{\INFPSI}_{\ell+1} = 
\frac{3(\ell+1)b_{\ell+1}}{(1-T)^{3\ell+4}} - %
\frac{2(\ell+1)b_{\ell+1} + (3\ell+2)c_{\ell+1}}{(1-T)^{3\ell+3}} +
\frac{(2\ell+1)c_{\ell+1}}{(1-T)^{3\ell+2}} \, .
\eeq
Remarking that $(1-T)\Vz \INF{\ell+1} + \left(\ell+1\right) \INF{\ell+1}
 = {\INFPSI}_{\ell+1}$, we have completed the proof of
$ \INF{\ell+1} \preceq W_{\ell+1}$.

\bibliographystyle{plain}

\end{document}